\documentclass[a4paper,aps,superscriptaddress,floatfix,nofootinbib,twocolumn,bibtex]{revtex4}

\usepackage[pdftex]{graphicx}
\usepackage{color}
\usepackage{amsmath,amssymb}
\usepackage{natbib}
\usepackage[english]{babel}

\begin{document}

\title{Analysis of bibliometric indicators for individual scholars
in a large data set}

\author{Filippo Radicchi}\email{f.radicchi@gmail.com}
\affiliation{Departament d'Enginyeria Quimica, Universitat Rovira i Virgili, Av. Paisos Catalans 26, 43007 Tarragona, Catalunya, Spain}

\author{Claudio Castellano}\email{claudio.castellano@roma1.infn.it}
\affiliation{Istituto dei Sistemi Complessi (ISC-CNR), Via dei Taurini 19, I-00185 Roma, Italy\\
Dipartimento di Fisica, “Sapienza” Universit\'a di Roma, P.le A. Moro 2, I-00185, Roma, Italy}

\begin{abstract}
Citation numbers and other quantities
derived from bibliographic databases are becoming
standard tools for the assessment of
productivity and impact of research activities. 
Though widely used, still their statistical properties
have not been well established so far.
This is especially true in the case of 
bibliometric indicators aimed at the evaluation
of individual scholars, because
large-scale data sets are typically difficult to be retrieved.
Here, we take advantage of a recently introduced large bibliographic
data set, Google Scholar Citations, which
collects the entire publication record of individual scholars.
We analyze the scientific profile of more than $30,000$
researchers, and study the relation 
between the $h$-index, the number of publications
and the number of citations of individual scientists.
While the number of publications of a scientist
has a rather weak relation with his/her $h$-index,
we find that the $h$-index of a scientist
is strongly correlated with the number
of citations that she/he has received so that
the number of citations can be effectively be used as 
a proxy of the $h$-index.
Allowing for the $h$-index to depend on both the number
of citations and the number of publications,
we find only a minor improvement.
\end{abstract}

\maketitle

\section{Introduction}
\label{intro}
Bibliographic databases play nowadays a 
crucial role in modern science. Citation numbers, or 
other measures derived from bibliographic data, 
are commonly used as quantitative 
indicators for the impact of research activities. 
Citation analysis has been criticized
~\citep{MacRoberts1989,MacRoberts1996,Adler2009}, 
and the true meaning of a citation 
can be very different from context to 
context~\citep{Bornmann2008, Hartley2012}. 
Despite these objections, the use of citations is widespread and
citation numbers are currently and frequently used for assessing
the impact of individual scholars~\citep{Hirsch2005, Egghe2006}, 
journals~\citep{Garfield2005}, departments~\citep{Davis1984}, 
universities and institutions~\citep{Kinney2007}. 
Especially at the level of individual scientists, numerical 
indicators based on citation counts are evaluation tools of 
fundamental importance for decisions about hiring~\citep{Bornmann2006} and/or 
grant awards~\citep{Bornmann2008a}.

\

\noindent Though widely used, numerical indicators
based on citation numbers are generally poorly understood~\citep{Lehmann2006}.
Even in the basic case of citation distributions
of papers, where data are easily collectable and analyzable,
there is no clear general picture. Depending on the study
performed and the data set analyzed, citation distributions
have been judged compatible with several possible
statistical distributions: power-law functions
~\citep{Price1965, Redner1998}, log-normal 
distributions~\citep{Stringer2008, Radicchi2008, Stringer2010}, 
stretched exponentials
~\citep{Laherrere1998, Wallace2008}, and others.
At the same time, however, while researchers have not yet
reached an agreement on the precise law governing
citation distributions, interesting properties
in citation data are nevertheless visible 
and detectable~\citep{Radicchi2012b}.
\\
Up to now, large-scale statistical analyses have been 
limited to the study of citations accumulated by papers
\citep{Wallace2008, Radicchi2012b}
or journals~\citep{Rosvall2008, West2010}.
In these cases data are easily collected from
the main bibliographic databases available on the market
and do not require special filtering procedures.
Conversely, the collection of bibliographic data
about individual scholars is much more
difficult. Simple searches on bibliographic databases
are generally unable to produce clean data sets because of evident
problems related to the proper disambiguation
of scientists. All studies conducted so far have 
been therefore limited either to small
sets of scientists~\citep{Hirsch2005, Bornmann2008a, 
Costas2008, BarIlan2008, Redner2010, Petersen2010c,
Petersen2010b, schriber2011, Petersen2011a, Petersen2012, Pratelli2012},
or to data sets subjected to
disambiguation problems~\citep{Lehmann2006, Radicchi2009, Cabanac2013}.
\\
Here, we take advantage of a data set
composed of more than $30,000$ [about 
two orders of magnitude larger than those used by~\citet{Hirsch2005, Bornmann2008a, 
Costas2008, BarIlan2008, Redner2010, Petersen2010c,
Petersen2010b, schriber2011, Petersen2011a, Petersen2012}]
individual scientific profiles.
The data set is rather clean
because profiles are directly managed by scientists themselves, who are
interested in providing correct information
about the outcome of their research activity.
We perform an initial exploratory analysis
of this data set, and show that the main basic quantities used in
research evaluation exercises obey well-defined statistical distributions.
We then use the data set to investigate (on a scale more than 10 times
larger than previous studies) the relation between the
$h$-index and other simple bibliometric indicators.

\section{Data set}
\label{sec:1}
We collected the scientific profiles of $89,786$ scientists
from Google Scholar Citations
(GSC, {\tt scholar.google.com/citations}) database.
The profile of each scholar reports the entire
publication record of the scientist, including the year of publication
and the number of citations accumulated by each publication according
to the Google Scholar database [for
studies about differences in the quantification of
bibliometric indicators of individual
scientists between Google Scholar and other popular
bibliographic databases see
~\cite{JACSO2005, JACSO2005B, meho2007, BarIlan2008, alonso2009}]. The profile is owned 
and managed by scientists themselves, who can delete and add
publications, even merge two publications if considered
initially as different in the database, and thus
provides a clean source of information. 
Scientists are requested to validate their profile by providing their
academic email address.
This validation ensures that the profile is actually managed by the
scientist. 
Finally, each scientist is required to provide a set of keywords which
identify the research fields in which the scientist is active.
\\
Data have been collected between June 29 and July 4, 2012. Number
of publications and citations reflect therefore the research activity
performed until that time.
We used 
an iterative procedure consisting in downloading the 
entire set of authors using an initial keyword (we used
``network science''), adding the other keywords used 
by these scientists, downloading the profile of new scientists
that are using these new keywords, and so on,
until we were able to discover neither new scientists nor new keywords. 
In total, we were able to identify $67,648$ different keywords
(see Fig.~1 for a word cloud of the most common keywords). 
It is important to notice that the database is in rapid
evolution and growth. For example, we used the same procedure
described above to download data in March 2012, and at that time the
data set was composed of $49,365$ scientists and $38,679$ keywords.

\

\noindent In order to be sure about the information provided by users, 
have a better control of the publication record of individual
scientists, and include only scholars with a sufficiently
long period of activity and sufficiently
large number of publications, we filter
the data set with the following restrictions:
\begin{enumerate}
\item We restrict our analysis
to the $83,897$ scientists who validated their
profile with an academic email.
\item We delete from the data
set publications that were published before year $1945$.
This was necessary in order to exclude from the data set
papers whose year of publication is wrong in GCS. Note also
that scientists with first publication before year $1945$,
if still active and with a validated profile,
would have academic ages longer than $67$ years.
\item We further restrict the attention
to scientists with at least $20$ publications 
and career length longer than or equal to $5$ years
(the academic age or career length is measured as the
difference between the publication years
of the first paper of a scholar and year $2012$). 
\end{enumerate}
In the rest of the paper,
we will present the result of the analysis based
on a total of $35,136$ scholars whose research
profile satisfies the aforementioned conditions.

\begin{figure*}[!ht]
\includegraphics[width=0.95\textwidth]{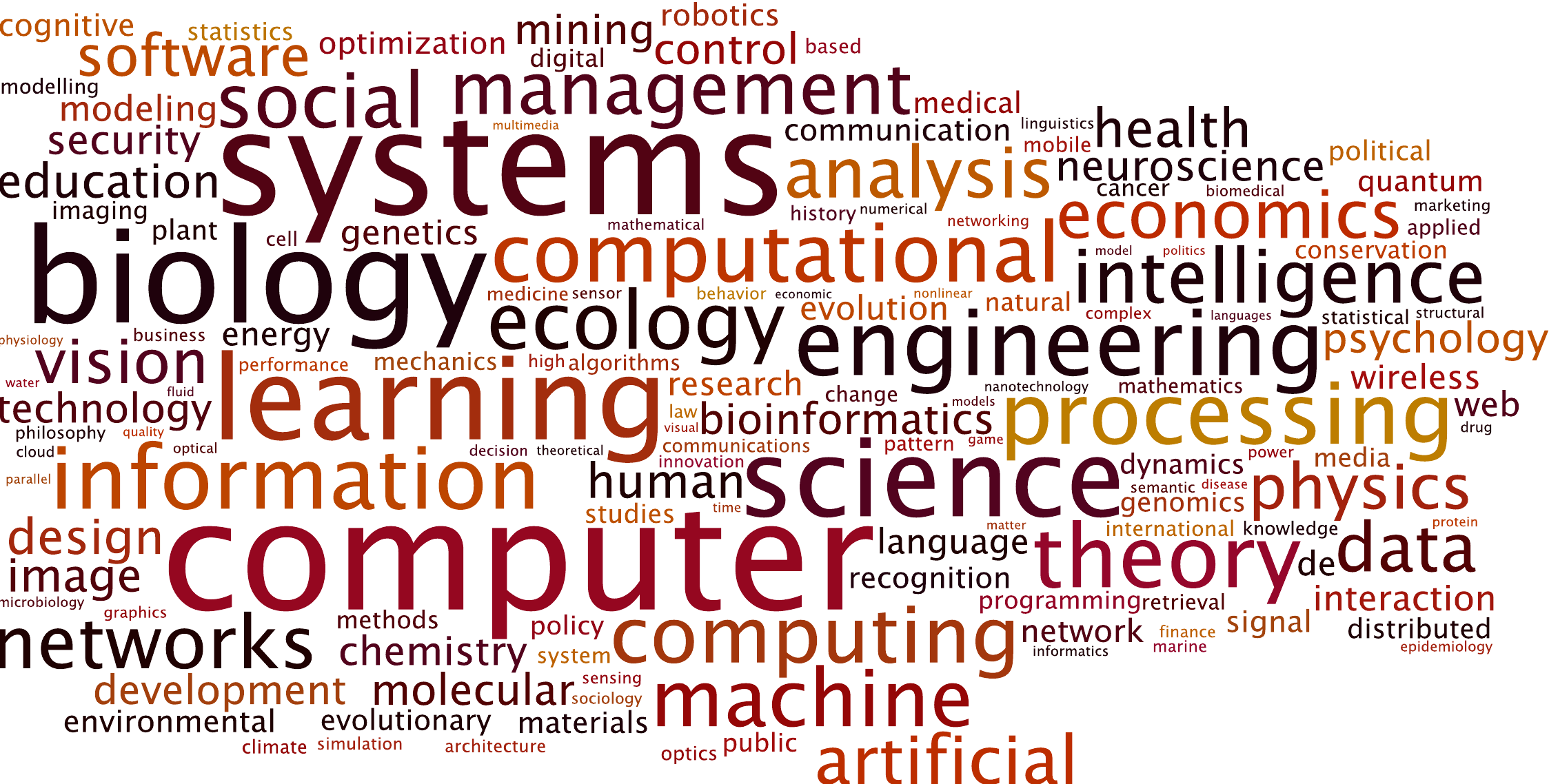}
\caption{Word cloud of the most common keywords
associated with the academic profiles in our data set.}
\label{fig:cloud}
\end{figure*}

\begin{figure*}[!ht]
\includegraphics[width=0.95\textwidth]{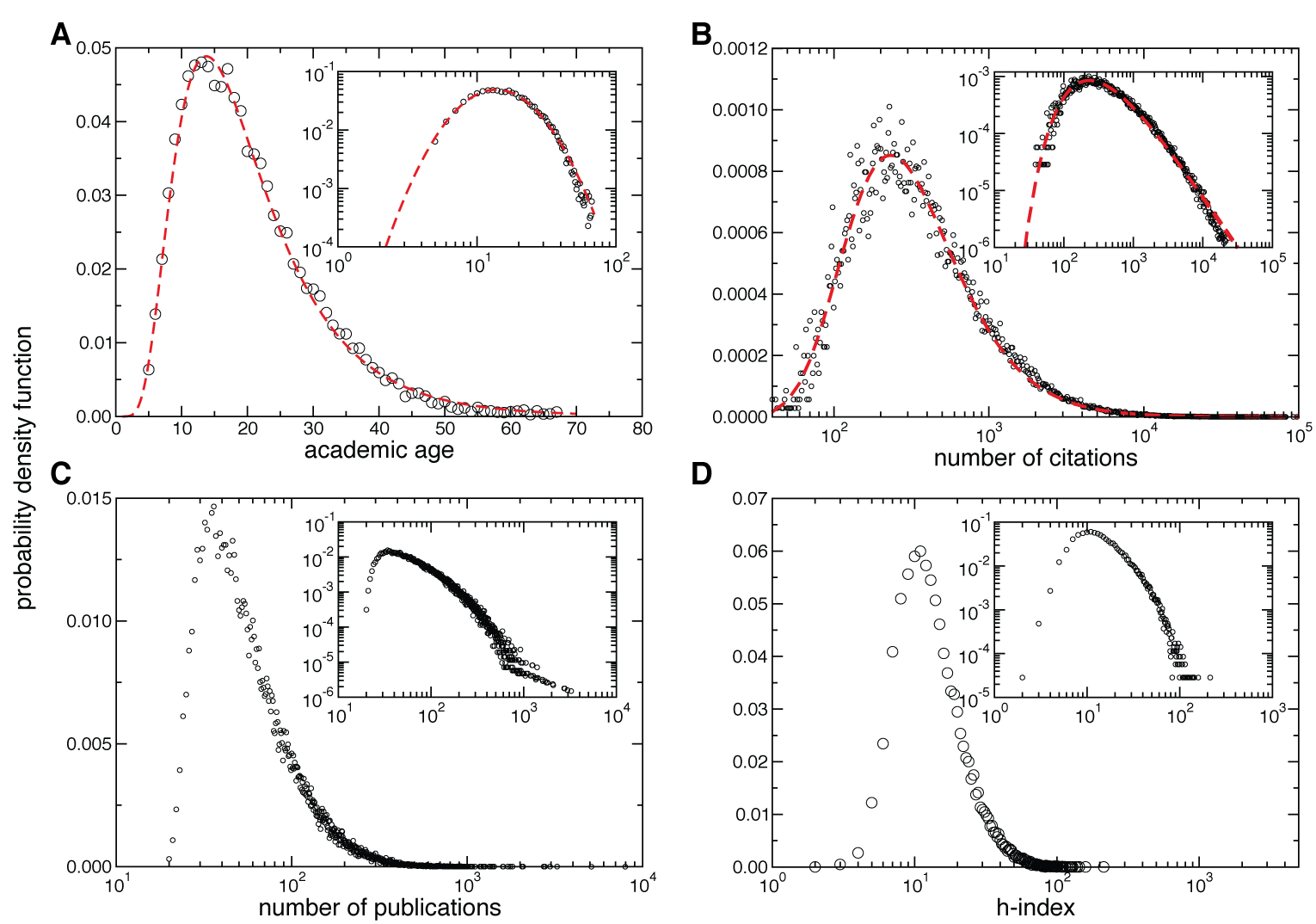}
\caption{{\bf A.} Probability density function of
the academic age of the scientists in the data set.
Data are fitted with a log-normal distribution [Eq.~\ref{eq:1}] with
parameters values $\hat{\mu}_A=2.89$ and $\hat{\sigma}_A=0.51$ 
(red dashed line).
{\bf B.} Probability density function of
the total number of citations received by the scientists in the data set.
Data are fitted with a log-Gumbel distribution [Eq.~\ref{eq:2}] with
parameters values $\hat{\nu}_C=6.42$ and
$\hat{\tau}_C=1.22$  (red dashed line).
{\bf C.} Probability density function of
the total number of publications produced by the scientists in the data set.
%Data are fitted with a log-Gumbel distribution with
%parameters values $\hat{\nu}_C=3.96$ and
%$\hat{\tau}_C=0.62$ (red dashed line).
{\bf D.} Probability density function of
the $h$-index of the scientists in the data set.
%Data are fitted with a log-Gumbel distribution with
%parameters values $\hat{\nu}_h=2.55$ and
%$\hat{\tau}_h=0.54$ (red dashed line).
}
\label{fig:1}
\end{figure*}

\section{Results}
\label{sec:2}
\subsection{General properties of the data set}

\noindent 
We first investigate general properties of the population in our data set.
In Fig.~\ref{fig:1}A, we show the composition of the population in
terms of academic age. The probability density function (pdf)
$P(A)$ of the career length
$A$ can be reasonably well described (by graphical inspection, although
the measured $p$-value does not support a good statistical compatibility)
by a log-normal distribution
\begin{equation}
P\left(A\right) = \frac{1}{A \, \sqrt{\pi \, \sigma^2} } \; e^{-\left[\log\left(A\right) - \mu \right]^2 / \left(2 \sigma^2\right)} \;\;,  
\label{eq:1}
\end{equation}
and the best estimate of parameters
(obtained with least square fit) of the
distribution are
$\hat{\mu}_A=2.89$ and $\hat{\sigma}_A=0.51$ (the
suffix $A$ is used to indicate that the parameters
have been calculated for the academic age $A$).
\\
The number of citations $C$ received by each scientist
($C$ is the sum of all citations accumulated by all papers
written by an author) is well fitted by (see Fig.~\ref{fig:1}B)
\begin{equation}
P\left(C\right) = \frac{1}{C \, \tau } \; e^{-z - e^{-z}} \;\;, 
\label{eq:2}
\end{equation}
where $z=\frac{\log\left(C\right) - \nu} {\tau}$, and the
best estimates of the parameters are $\hat{\nu}_C=6.42$ and
$\hat{\tau}_C=1.22$.  Eq.~\ref{eq:2} is a generalization to the logarithms
of the well-known Gumbel function that usually
appears in the description of the statistics
of extreme values. 

%The same type of distribution is also
%valid for the total number of publications $N$  ($\hat{\nu}_C=3.96$ and
%$\hat{\tau}_C=0.62$, Fig.~\ref{fig:1}C) and 
% the $h$-index  ($\hat{\nu}_h=2.55$ and
%$\hat{\tau}_h=0.54$, Fig.~\ref{fig:1}D).
In Figs.~\ref{fig:1}C and D, we report the
pdfs of the number of publications $N$
and the $h$-index, respectively. 
In these cases, we tried to fit
the distributions with both Eqs.~\ref{eq:1}
and~\ref{eq:2}, but none of them was
able to describe entirely the pdfs obtained with
data. It is, however, interesting
to note that the pdf of the number of publications
per author is neither a strict decreasing function
nor a power-law function
as often assumed in the literature~\citep{Egghe2010},
but instead the pdf in Fig.~\ref{fig:1}C shows a clear
peak and a decay faster than a power-law at large
values of $N$.
\\
In general, the results presented in Fig.~\ref{fig:1}
depend on the choices we made in the selection
of the authors. For example, the peak position 
of the $P\left(A\right)$ [Fig.~\ref{fig:1}A] moves
from $A=14$ to $A=6$ if we remove the restriction
on the minimal number of publications needed to enter
in the sample. Similar considerations are also
valid for the other pdfs. On the other hand, our
choices do not affect the tail of the pdfs, and more
generally their shapes. For example, even if the peak moves to lower values
when we include all authors in the data set,
the pdf of Fig.~\ref{fig:1}A still can be
reasonably well described by a log-normal distribution.

%\\ 

\begin{figure*}[!htb]
  \includegraphics[width=0.95\textwidth]{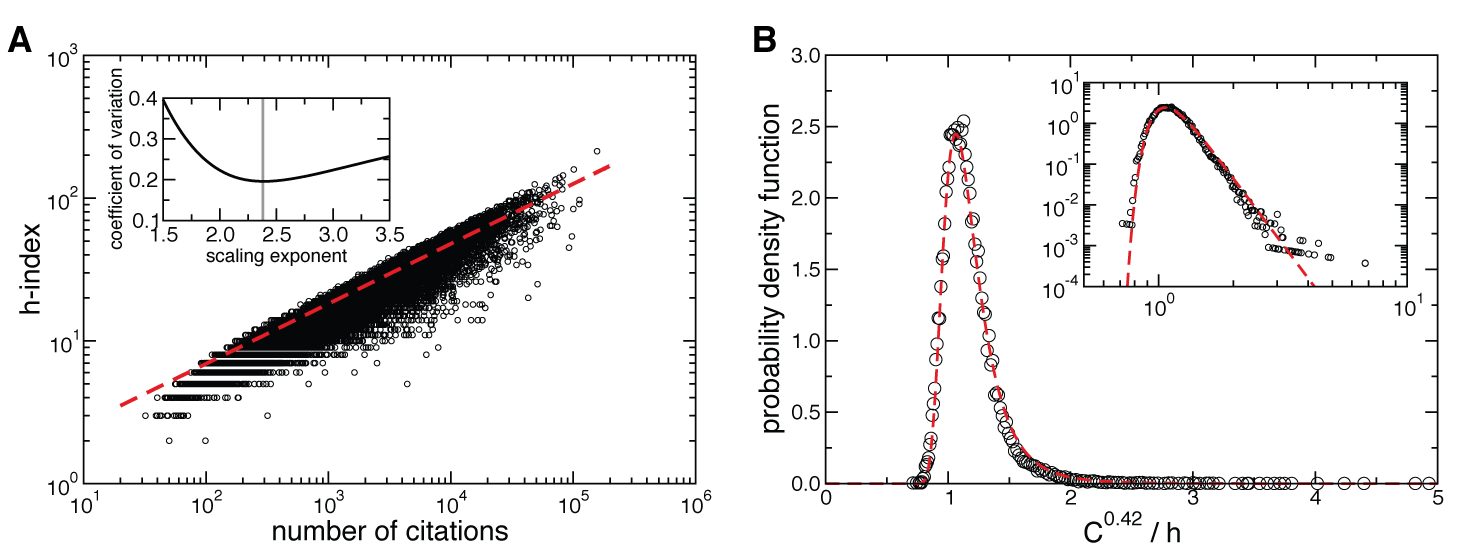}
\caption{{\bf A}. Relation between the $h$-index
and the number of citations $C$ for each scientist
in our data set. We fit data with a
power-law function [Eq.~\ref{h_vs_C}], whose best
estimate of the exponent equals  $1/\hat{\alpha}_{h, C} 
= 1/2.39 = 0.42$ (dashed line).
{\bf B}. Probability density function of the quantity
$C^{0.42}/ h$. This function is fitted by
a log-Gumbel distribution [Eq.~\ref{eq:2}] with parameter values
$\hat{\nu}_{h,C} =0.08$ and $\hat{\tau}_{h,C} =0.14$
(dashed line). The inset shows the same
as the main plot but in a double-logarithmic scale.}
\label{fig:2}
\end{figure*}

\begin{figure*}[!htb]
  \includegraphics[width=0.95\textwidth]{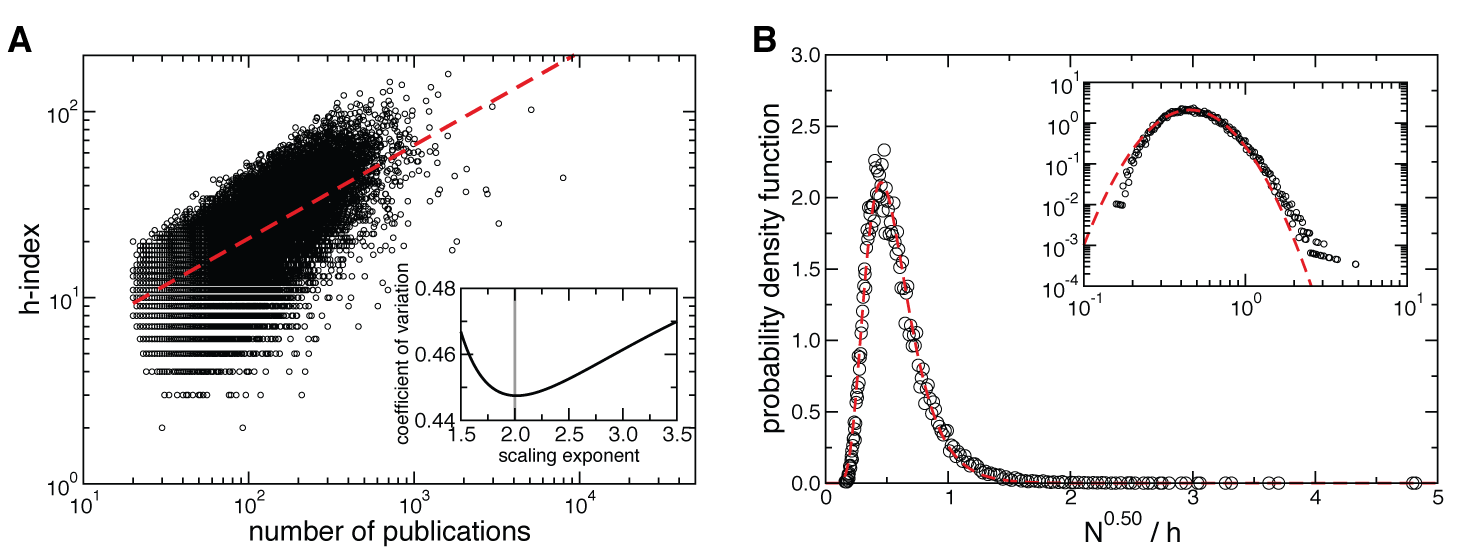}
\caption{{\bf A}. Relation between the $h$-index
and the number of publications $N$ for each scientist
in our data set. We fit data with a power-law function, whose best
estimate of the exponent equals  
$\hat{\alpha}_{h, N} = 0.50$ (dashed line).
{\bf B}. 
Probability density function of the quantity
$N^{0.50}/h$. This function is fitted by
a log-normal distribution with parameter values
$\hat{\mu}_{h,N} =-0.64$ and $\hat{\sigma}_{h,N}=0.39$ 
(dashed line). The inset shows the same
as the main plot but in a linear-logarithmic scale.}
\label{fig:3}  
\end{figure*}

\begin{figure*}[!htb]
  \includegraphics[width=0.95\textwidth]{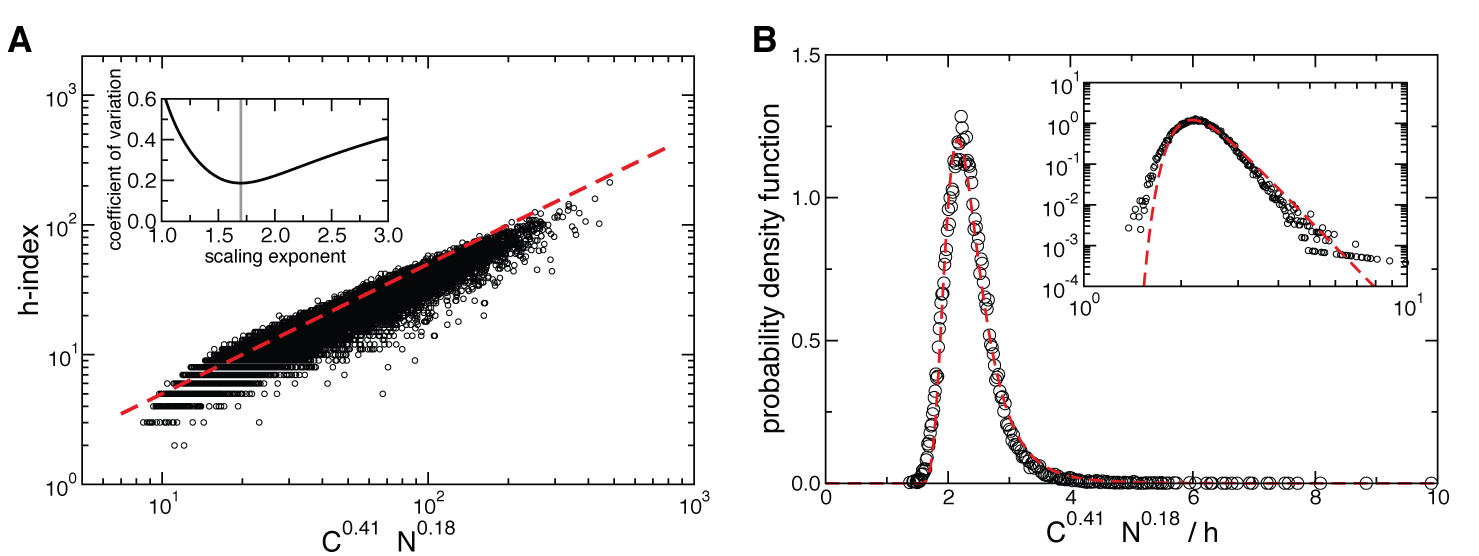}
 \caption{{\bf A.} Relation between the $h$-index,
the number of citations $C$
and the number of publications $N$ for each scientist
in our data set [Eq.~\ref{eq:final}]. 
The best estimate of the exponent equals  
$\hat{\alpha}_{h, C, N} = 1.70$ (see inset).
Effectively, when $h$ is plotted against $C^{0.41} \, N^{0.18}$ 
(i.e., Eq.~\ref{eq:final} for
$\alpha_{h, C, N} = \hat{\alpha}_{h, C, N}$),
we recover a good linear behavior (dashed line).
Probability density function of the quantity
$C^{0.41} N^{0.18} / h$. This function is fitted by
a log-Gumbel distribution with parameter values
$\hat{\nu}_{h,C,N} =0.79$ and $\hat{\tau}_{h,C,N} =0.14$
(dashed line). The inset shows the same
as the main plot but in a double-logarithmic scale.}
\label{fig:4}  
\end{figure*}

\subsection{Relation between the $h$-index and other indicators.}

In this subsection we test empirically some relations between the
$h$-index and other bibliometric indicators, which have been proposed
in the literature.

Already in his original paper, Hirsch himself presented a very
simple model for the accumulation of citations, which implies
a correlation between the $h$-index and the total number $C$ of citations
received by an individual
\begin{equation}
h \sim C^{1/\alpha_{h,C}}
\label{h_vs_C}
\end{equation}
with exponent $\alpha_{h,C}=2$.
A correlation of this type between  $h$ and $C$ has been verified empirically
for small data sets~\citep{VanRaan2006,Redner2010,Spruit2012}.
In Fig.~\ref{fig:2}A we plot, for each author in our data set, 
the $h$-index {\it vs.}
the number of citations accrued $C$, in log-log scale.
The correlation is rather strong (linear correlation
coefficient measured in log-log scale $R_{h, C} = 0.95$) %0.82$)
supporting the hypothesis of
a scaling relationship between these two quantities.
Since we are interested in finding the scaling
between $h$ and $C$ that would provide the strongest
relation between them, we determine the best estimate of
$\alpha_{h, C}$
as the one that produces the most localized distribution
of the ratio $ x = C^{1/\alpha_{h,C}} / h$. We quantify
the localization of the distribution of $x$ by means
of the so-called coefficient of variation $\sigma_x /\langle x \rangle$,
i.e., the ratio between the standard deviation and the average
value of $x$~\citep{Hendricks1936, Costas2007}.
The best estimate of the power-law exponent in Eq.~\ref{h_vs_C}
is obtained as the value of $\alpha_{h,C}$ that minimizes
the coefficient of variation. This way, we find 
$1/\hat{\alpha}_{h, C} =1/2.39 = 0.42$ 
(see Fig.~\ref{fig:2}A), which is quite close to Hirsch's original prediction.
As additional fitting procedures, we also
calculated the best estimate of $\alpha_{h,C}$
as the one minimizing the kurtorsis  of the distribution of $x$,
or the one minimizing the mean square displacement
$\chi^2 = \sum_i \left(h_i - a_{h,C} \, C^{1/\alpha_{h,C}} \right)^2$,
where the sum runs over all
authors in the data set and $ a_{h,C}$ is an additional fitting parameter.
In all cases, we find similar values for the best estimate 
of $\alpha_{h,C}$. 

To further characterize the relation
between $h$ and $C$, we study the statistical properties
of the quantity $C^{1/\hat{\alpha}_{h,C}} / h$ in Fig.~\ref{fig:2}B.
We find that the distribution is narrowly peaked
around a value close to 1, and that can be
nicely fitted by the log-Gumbel distribution
of Eq.~\ref{eq:2} whose best parameter
estimates are $\hat{\nu}_{h,C} =0.08$
and $\hat{\tau}_{h,C} = 0.14$.

The relation reported in Eq.~\ref{h_vs_C} between $h$ and $C$ can be easily
derived~\citep{Egghe2010} assuming that the distribution of the
number of citations accrued by the publications of a single author
is a power-law $f(c) = K c^{-\alpha}$ ($K$ being a normalization constant).
The exponent $1/\alpha \lesssim 1/2$ found numerically implies a
distribution $f(c)$ decaying with an exponent slightly larger than 2.
If the distribution is a perfect power-law and $\alpha>2$,
the same power-law relation also holds between $h$ and the total
number of publications $N$~\citep{Egghe2006c,Glaenzel2006}:
$h \sim N^{1/\alpha}$.
Since the distribution $f(c)$ is not a perfect power-law over the whole
range of $c$ values, it is worth checking empirically the validity 
of such a power-law relationship, by allowing the scaling exponent
to be possibly different from $1/\alpha_{h, C}$
\begin{equation}
h \sim N^{1/\alpha_{h, N}}.
\label{h_vs_N}
\end{equation}
In Fig.~\ref{fig:3}A we find that $h$ and $N$ are
correlated ($R_{h,N} = 0.72$), %0.63$), 
although less than in the previous
case. The best estimate for the exponent, again obtained by minimizing
the coefficient of variation, is $\hat{\alpha}_{h, N} =2.0$.
Notice that this value indicates that $h$ depends differently
on $C$ and on $N$, thus contradicting the hypothesis that $f(c)$ is,
for all scholars, a pure power-law over its whole range.
In Fig.~\ref{fig:3}B we plot the pdf of the quantity $N^{0.50}/h$,
which is approximately fitted by a log-normal distribution with parameters
values $\hat{\mu}_{h, N} = -0.64$ and $\hat{\sigma}_{h, N} = 0.39$.
Also in this case we find quite a narrow distribution, but the
value of the coefficient of variation indicates a worse agreement
with data with respect to one found for $h$ {\it vs.} $C$. 
Indeed, the inset of Fig.~\ref{fig:2}A shows that the minimum coefficient
of variation (corresponding to $\hat{\alpha}_{h, C} =2.39$) is around 0.19,
while the minimum in the inset of Fig.~\ref{fig:3}A is around 0.45.
\\
Under the power-law assumption for $f(c)$, it is also possible
to express $h$ as a function 
of both $N$ and the average number of citations per paper $\chi=C/N$, 
obtaining~\citep{Schubert2007,Iglesias2007}
\begin{eqnarray}
h &\sim & \chi^{(\alpha_{h, C, N}-1)/\alpha_{h, C, N}} \, N^{1/\alpha_{h, C, N}} \\ 
& \sim & C^{(\alpha_{h, C, N}-1)/\alpha_{h, C, N}} \,  N^{(2-\alpha_{h, C, N})/\alpha_{h, C, N}} \, .
\label{eq:final}
\end{eqnarray} 
Minimizing the coefficient of variation for the quantity 
$C^{(\alpha_{h, C, N}-1)/\alpha_{h, C, N}} \,  N^{(2-\alpha_{h, C, N})/\alpha_{h, C, N}} /h $,
we determine the best estimate of $\alpha_{h, C, N}$
as $\hat{\alpha}_{h, C, N} = 1.70$ (inset of Fig.~\ref{fig:4}A),
implying $h \sim C^{0.41} \, N^{0.18}$.
In the main panel of Fig.~\ref{fig:4}A we plot, for each author, 
$h$ {\it vs.} $C^{0.41} \, N^{0.18}$
finding a good agreement with the expected linear behavior.
Also in this case we find that the quantities are correlated 
($R_{h,C \cdot N} = 0.94$).
The pdf of the rescaled index $C^{0.41} \, N^{0.18} / h$ is
well peaked (Fig.~\ref{fig:4}B), and the distribution
can be reasonably well fitted by a log-Gumbel distribution
with best-fit parameters
$\hat{\nu}_{h, C, N} = 0.79$ and $\hat{\tau}_{h, C, N} = 0.14$.

As the relation $h \sim C^{0.41} N^{0.18}$ shows,
the dependence of $h$ on $C$ is much stronger than the one on $N$. 
The comparison between the minimum coefficients of variation measured 
for the three scaling assumptions
(Eqs.~\ref{h_vs_C}, ~\ref{h_vs_N} and ~\ref{eq:final}) 
indicates that allowing for a dependence on both $C$ and $N$
leads only to a marginal improvement over considering only the
dependence on $C$ (the coefficient of variation changes only from
$0.19$ to $0.18$) while a dependence only on $N$ performs definitely worse.
The presence of a term dependent on $N$ in Eq.~\ref{eq:final} brings
only a little improvement and leaves the exponent of the dependence
on $C$ practically unaltered ($0.41$ {\it vs.} $0.42$).

\section{Conclusions}
\label{sec:4}

Statistical analysis of bibliometric indicators
devoted to the evaluation of individual scholars
is usually difficult because of the lack
of large and clean data sets describing accurately
the publication records of researchers. This is
a general problem that regards every bibliographic database
available on the market, and is also the main
reason for which the studies performed so far
on the characterization of the bibliographic profile
of individual scientists have been rarely based
on more than $1,000$ individuals.
In recent years, however, some main
bibliographic databases have started
to allow individual scientists to freely manage
their publication profiles with specially
designed on-line tools.
This is the case of the recently created
ResearcherID by Thomson Reuters ({\tt http://www.researcherid.com}),
Mendeley profiles ({\tt http://www.mendeley.com})
and also of Google Scholar Citations \\
({\tt scholar.google.com/citations}).
In all these online administration systems, scholars
manage directly their profiles by adding,
deleting and correcting their publication records, and
thus the information provided can be
considered accurate because it is in the interest of 
researchers to provide an accurate and up-to-date source of
information regarding their research production.
\\
Here, we took advantage of the entire data set
of Google Scholar Citations as of June 29, 2012.
The data set is composed of more than $30,000$ individual
scholars working in research institutions worldwide.
Although Google Scholar Citations represents
a relatively clean set of data
describing the academic records
of individual researchers, it is important to stress that 
our set of data does not represent a random sample
of researchers because the presence
of a researcher in the system is
subjected to various types of factors that, as
a matter of fact, bias the sample. 
First, our data set is mainly composed of
relatively young scientists (Fig.~2A) who are
able to create a profile, validate it with their
email, and manage it. Second, the profiles that
compose the data set are certainly those
of scholars who want to promote their research
with the use of modern information-technology
tools. Finally, although the scientists present in our data set
have various fields of expertise, some scientific disciplines 
are clearly over-represented and others underrepresented (see Fig.~1).
We would like to further emphasize that
the entire data set analyzed here is clearly subjected
to all the limitations of the Google Scholar
database (eventual presence
of fake publications, duplication of citations, etc.) 
that have been deeply studied in the
literature~\citep{Harzing2008,JACSO2010,Labbe2011}.
\\
\noindent Taking into account
the formerly mentioned
limitations of our data set, here  
we provided an exploratory analysis on some basic statistics
for single authors and focus in some detail on the
relation between the $h$-index, the number of publications
and the number of citations of individual scientists.
We found three main results:
\begin{enumerate}

\item The $h$-index $h$ is strongly correlated
with the total number of citations $C$ received by
a scientist with his/her own scientific production.
In particular, we find $h \sim C^{0.42}$
in qualitative agreement with the early hypothesis by
Hirsch~\citep{Hirsch2005} validated empirically on small data
sets~\citep{VanRaan2006,Redner2010,Spruit2012}.

\item The $h$-index is also shown to be correlated with the
number of publications $N$, but this relation is much less precise
than the one observed for $C$. 

\item It is possible to combine both
dependencies into a single power-law relation
$h \sim C^{0.41} \, N^{0.18}$. This law, however,
provides only a slight improvement with respect
to the power-law relation that connects only $h$ and $C$.

\end{enumerate}

Our results represent a large-scale validation
of formerly postulated conjectures.
While the exact values of the measured
power-law exponents might be data set dependent, we believe
that the main message has a validity that goes beyond the data
analyzed here: the total number of citations $C$ received 
by a scientist can be used as a effective proxy of his/her $h$-index.

The fact that $h$ is strongly correlated with $C$ and much more
weakly with the total number of publications $N$ is evidence that
the distribution $f(c)$ of citations accrued by publications of
a single researcher is not a pure power-law over its whole range.
Please note that we do not exclude the possibility
that this fact
is a consequence of the
sample used in our analysis, where
scholars have academic age typically shorter
than the one of an average researcher and thus
individual citation distributions may have not yet
reached as sufficient level of stationarity.
Our results, however, call for more work to better characterize and understand
the activity and citations profile of individual scholars, their
common features and their variations.
The large data set provided by Google Scholar Citations constitutes
an ideal tool for this endeavor. The present study represents only
a first attempt to scratch the surface of such a treasure trove.

\end{document}